# Building Multilingual TTS using Cross-Lingual Voice Conversion

*Qinghua Sun, Kenji Nagamatsu*

Research & Development Group, Hitachi, Ltd., Tokyo, Japan.
{qinghua.sun.ap, kenji.nagamatsu.dm}@hitachi.com

## Abstract

In this paper we propose a new cross-lingual Voice Conversion (VC) approach which can generate all speech parameters (MCEP, LF0, BAP) from one DNN model using PPGs (Phonetic PosteriorGrams) extracted from inputted speech using several ASR acoustic models. Using the proposed VC method, we tried three different approaches to build a multilingual TTS system without recording a multilingual speech corpus. A listening test was carried out to evaluate both speech quality (naturalness) and voice similarity between converted speech and target speech. The results show that *Approach 1* achieved the highest level of naturalness (3.28 MOS on a 5-point scale) and similarity (2.77 MOS).

## 1. Introduction

We are building a communication robot with an emphasis on communication with humans. At first, only Japanese was supported, but we aimed to expand it into multilingual support for English, Chinese, and Korean, which are commonly supported in the public spaces of Japan. It would be felt unnatural for the same robot to change its voice for each of the multiple languages, so we aimed multilingual TTS with a single speaker.

In order to realize conventional multilingual TTS, it is necessary to record a multilingual speech corpus using a single speaker, but the difficulty of recording increases in accordance with the number of languages. When further increase is considered, multilingual TTS supporting 3 or more languages is essentially impossible using conventional methods. In addition, it is desirable to expand multilingual support in a way that preserves the character of the generated speech. This is because we don't want to change the voice of our robot which was designed with the characteristics of the robot in mind and has been used for a long time.

So, our Multilingual TTS aims to build a system which can synthesize speech in a specific language not spoken by the target speaker. There are several approaches to achieve this goal. One is the GMM-HMM-based method [1, 2, 3] which trains two independent HMM-based TTS systems from data recorded by a bilingual speaker, which are then used in a framework in which the HMM states are shared across the two languages in decision tree-based clustering. A state mapping process is used to obtain mapping information and then apply it to the target speaker to synthesize speech in the target language. Then, using methods such as Variational Autoencoder, Adversarial Generative Networks (GAN), cross-lingual VC also can achieve a multilingual TTS without a *real* multilingual speech corpus. In this research, we focus on the VC approach and tried to find a versatile way to build Multilingual TTS systems.

VC is a technique to modify the speech from speaker A (source) to make it sound like speech from speaker B (target) without changing linguistic information [4]. Depending on whether the source speaker and target speaker speak the same language, VC can be broadly divided into intralingual VC and cross-lingual VC. For intralingual VC, most existing methods rely on parallel data during training, where it is necessary for the source and target speakers to record the same utterances[5]. But for cross-lingual VC (including our case), the source and target speakers speak different languages [6, 7]. Hence, parallel data is not available. To achieve cross-lingual conversions, several non-parallel VC techniques like Variational Autoencoder [8, 9] and Adversarial Generative Networks (GAN) [10, 11, 12] have been proposed. However, there is still a big gap between converted speech and natural speech even in intralingual VC. Recently, a new non-parallel VC approach using PPG, which represents the linguistic information of speech data, shows high performance in this domain [13]. PPG-based cross-lingual VC between English and Mandarin speakers has been reported in [14], which utilizes an English ASR system to obtain monolingual PPG in English. During training, English PPG is trained to be mapped to Mel-cepstrums (MCEPs) extracted from Mandarin utterances of target speaker. During testing, using the same English ASR system, PPGs are is extracted from a generated English speech. Then, these PPGs are passed to the trained model to obtain the converted MCEPs. Recently, English PPG was expanded bilingual PPG [15]. We got a great hint from this paper and extend monolingual PPG to clustered-state-based PPG to covert all speech parameters (not only acoustic features but also prosodic features) at the same time. We hope this can help us to improve both speech quality and speaker similarity to the target speaker.

In this paper, we build a multilingual (Japanese, Chinese and English) DNN-based TTS, in which speech is generated for all the languages with the same voice, a Japanese female speaker who can not speak Chinese or English at all.

## 2. VC using multi-lingual PPGs

Most of the PPG-based VC systems covert acoustic features and prosodic features independently [13,14,15]. But in our experience, when acoustic features and prosodic features are generated (or modified) independently, mismatch of these features will cause a declination in voice quality of generated speech [16]. In addition, we observed some deteriorations when F0 extraction errors occurred in a conventional VC system. These phenomena are often observed at voiced consonants or the tail of the sentence where vocal-fold vibration is not stable. F0 extraction mistakes in speech recorded in realistic (noisy) environments were especially noticeable. It is clear that these F0 extraction mistakes greatly affect the voice quality of converted speech.

In order to achieve more stable voice quality, we devised a method to simultaneously predict all voice parameters, including F0 patterns, without extracting F0 directly from speech. Specifically, this method simultaneously predicts 3 voice parameters (MCEP, LF0 and BAP) after extracting the PPGs from the inputted speech.

In our kaldi-based ASR, each dimension of the DNN output layer corresponds to an HMM state obtained using phoneme clustering. In our VC method, we used vectors of DNN output (around 5000 - 6000 dimensions) directly instead of the *real PPGs* which only have around 100 - 300 dimensions.

Most of the PPG-based VC systems convert prosodic features using a linear conversion approach, because it is supposed that PPGs have a weak relationship with prosodic features. As our observation, these DNN outputs (below we will call PPG) have more information and higher relationship with prosodic features than real PPGs. Furthermore, the prosodic features can be represented well if we combine Japanese, Chinese and English PPGs together.

Fig. 1 is a rough explanation of how the proposed method is carried out.

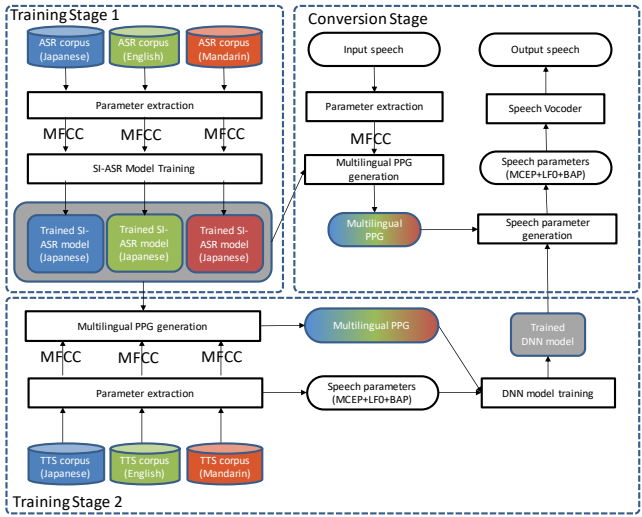

Fig. 1: Schematic diagram of proposed VC method using multilingual PPGs

In Training Stage 1, SI-ASR (Speaker-Independent Automatic Speech Recognition) acoustic models of Japanese, Chinese, and English were respectively trained from their own speech corpus for ASR. In particular, we hypothesized that F0 movement over a wide area (the range spans multiple syllables) can be represented by the PPG of Japanese, which is a typical accent language and the local (within syllables) F0 movement can be represented by the PPG of Chinese, which is a typical tonal language.

In Training Stage 2, multilingual PPGs are generated by merging the Japanese, Chinese, and English PPGs extracted from a TTS corpus using the SI-ASR models for each language trained in Training Stage 1. Concurrently, speech parameters are extracted from a TTS speech corpus. Then DNN models are trained using multilingual PPG as input vectors and speech parameters as output vectors.

In the Conversion Stage, we use the acoustic model for each language trained in Training Stage 1 to generate multilingual PPGs from input speech. Then, we use the DNN model trained in Training Stage 2 to predict the speech parameters from the generated multilingual PPGs. Finally, we use a speech vocoder to synthesize speech from the speech parameters. Considering the computational complexity of neural vocoders such as WaveNet, in this research we used a classical vocoder developed by Hitachi (Undisclosed) for our experiments.

## 3. Building multi-language TTS using proposed VC

Recently, deep learning-based TTS systems have been studied intensively, and some recent studies are achieving surprisingly clear results. The TTS systems based on deep neural networks include Zen's work in 2013 [17], the studies based on DNN [18, 19, 20], and end-to-end speech synthesis system based on deep learning has made great progress such as Tacotron [21], Tacotron2 [22], DeepVoice3 [23], ClariNet [24], Char2wav [25] and VoiceLoop [26]. Although the end-to-end TTS can generate natural speech which is close to humans, these models require a large amount of speech data from one speaker to obtain good quality. But traditional text-to-speech (TTS) synthesis methods still have most of the TTS market share. Considering these situations, our target is DNN-based classical TTS, in which an NLP module is developed by Hitachi, a Parameter generation module is implemented based on Merlin [27], and our original classical real-time vocoder will be used in waveform generation modules.

In this research we proposed 3 approaches to build multilingual (Japanese, English and Chinese) TTS systems using the proposed VC method.

### 3.1. Approach 1: Covert speech corpus using VC

A multilingual speech synthesis system can be constructed using a multilingual speech corpus with the voice quality of a single speaker. However, recording speech in multiple languages from a single speaker is extremely difficult, and recording speech in 3 or more languages from a single speaker is essentially impossible. The simplest method is to convert speech corpora recorded from different speakers for each language to a single speaker's voice quality as shown in Fig. 2, then build a multilingual TTS system normally.

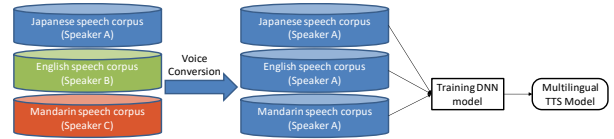

Fig. 2: Schematic diagram of approach1 (making multilingual speech corpus by VC)

### 3.2. Approach 2: Training model using GAN and VC

GAN (Generative Adversarial Network) is one generation method which consists of a generator which generates speech, and a discriminator which distinguishes the difference between generated speech and given speech (which we call "target speech" in our research). As Fig. 3 shows, multilingual DNN model can be trained by the GAN approach, in which the generator is trained using Chinese, Japanese, and English speech corpora (with different speakers), and the discriminator is given the Japanese speaker's speech data as the target speech.

Ideally, the discriminator is only trained to identify if the speech is from the target speaker or not. But in fact, not only speakers but also languages are different in the training data for the discriminator. Therefore, the discriminator tends to distinguish F0 movement similar to Japanese or English. Sometimes the inflection at the ending of a normal calm sentence generated in English may rise, while in Japanese the F0 at the end of sentence may stay high even in a calm sentence. Therefore, when a Japanese speech corpus was the target speech and GAN was used for training, the Japanese F0 characteristics were trained, resulting in unnatural generated English prosody.

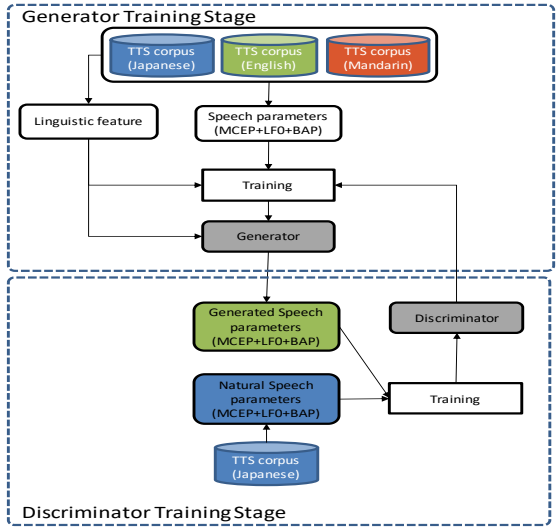

Fig. 3: Schematic diagram of Approach 2 using original Japanese TTS speech corpus in Discriminator Training Stage

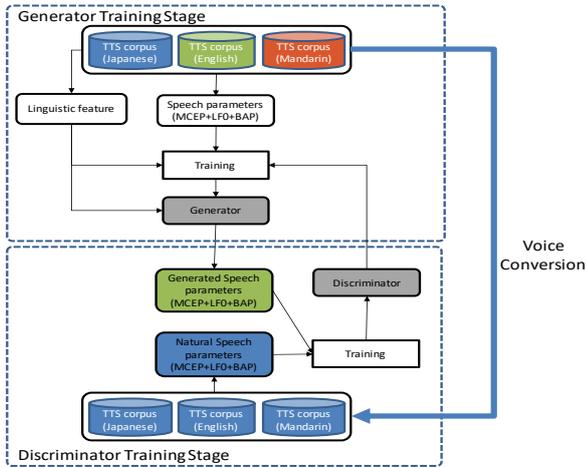

Fig. 4: Schematic diagram of proposed Approach 2 using converted multilingual TTS speech corpora in the Discriminator Training Stage.

In our research, we improved the training step by using VC. This proposed method achieved better results by replacing the Japanese speech corpus with the multilingual speech corpus using voice quality conversion, as shown in Fig. 4.

### 3.3. Approach 3: Convert synthesized speech using VC

To support some existing TTS systems which can not be retrained or built using the unit-selection method, we also propose an approach that converts generated speech using VC as a post process. In this research, we first build a multilingual TTS engine from a multilingual speech corpus with multiple speakers in the Training Stage. Then in the generation stage, generated speech is converted to the target speaker's voice using our proposed VC method.

This approach is extremely flexible and can be expected to apply to any speech synthesis system, but because of the huge computational complexity required for the VC process, a resource capable of computation such as a GPU is required to realize real-time processing, so productization is thought to be difficult at this point.

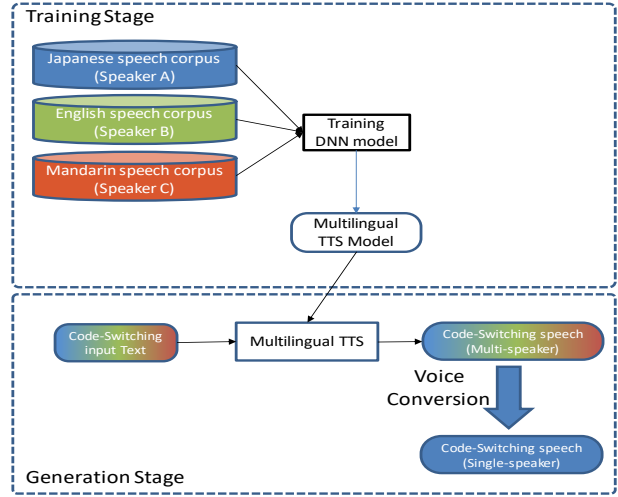

Fig. 5: Schematic diagram of Approach 3

## 4. Experiment

Here we performed speech synthesis experiments in order to verify an appropriate approach to realizing a multilingual synthesis system.

Considering the difficulty of evaluating the sound quality of multiple languages, in our research we used the voice quality of a Japanese speaker (named NKY) to construct each system and performed an evaluation of synthesized English speech only. (ELN is the English speaker in our English speech corpus.)

The conditions of the experiment are as follows:

**Listening environment:**
All speech was listened to using headphones in a quiet room.
**Presentation method:**
All speech was presented randomly.
**Evaluators:**
Naturalness was evaluated by 1 American native English speaker.
Similarity was evaluated by 4 native Japanese speakers.
**Input text:**
30 sentences of English news in addition to a training corpus were used as the input text.
**Speech samples for evaluation:**
Speech samples generated from 7 TTS systems (Table1, 2) and recorded speech were used in the listening test.
Table 1 shows the details of each speech sample. Table 2 shows the details of the DNN architectures used when generating each speech sample. Table 3 shows the details of the ASR models used in this experiment.
**Evaluation method:**
The method used to evaluate similarity was to listen to the generated speech, compare it to the reference speech (the Japanese speaker NKY's recorded speech (Speech A) and the English speaker ELN's recorded speech (Speech B)), and select one from the following four options.

In brief, the higher the points, the closer the generated speech is to the voice quality of the Japanese speaker NKY.
1: This was said by Speaker B (confident).
2: This was said by Speaker B (not confident).
3: This was said by Speaker A (not confident).
4: This was said by Speaker B (confident).

To evaluate naturalness, a subjective evaluation was performed using 5-scale MOS (5:Excellent, 4:Good, 3:Fair, 2:Poor, 1:Bad).

Table 1: Details of each speech samples.

|  | Whose voice | Approach |
|---|---|---|
| ELN_recording | ELN | Recording |
| ELNen_DNN | ELN | DNN model trained by EEV*1 |
| ELNen_LSTM | ELN | DNN model trained by EEV |
| NKYen_VC | NKY | Approach 3 (Fig. 5) DNN model trained by JNV*2 |
| NKYen_DNN | NKY | Approach 1 (Fig. 2) DNN model trained by ENV*3 |
| NKYen_GAN_NKYen | NKY | Approach 2 (Fig. 3) Using EEV for training generator and ENV for training discriminator |
| NKYen_GAN_NKYja | NKY | Approach 2 (Fig. 4) Using EEV for training generator and JNV for training discriminator |
| NKYen_LSTM | NKY | Approach 1 (Fig. 2) DNN model trained by ENV |

*1: EEV means English speech corpus with ELN's Voice.
*2: JNV means Japanese speech corpus with NKY's Voice.
*3: ENV means English speech corpus converted to NKY's Voice.

Table 2: DNN architectures used in current experiment.

|  | DNN architectures | |
|---|---|---|
|  | Type | Hidden layer size |
| ELN_recording | None | None |
| ELNen_DNN | Feed-forward | 1024×6 layer |
| ELNen_LSTM | Bi-LSTM | 256×4 layer |
| NKYen_VC | Feed-forward | 1024×6 layer |
| NKYen_DNN | Feed-forward | 1024×6 layer |
| NKYen_GAN_NKYen | Bi-LSTM | 256×4 layer |
| NKYen_GAN_NKYja | Bi-LSTM | 256×4 layer |
| NKYen_LSTM | Bi-LSTM | 256×4 layer |

Table 3: details of ASR model used in current experiment.

|  | Japanese | English | Mandarin Chinese |
|---|---|---|---|
| Frame shift | 10ms | 10ms | 10ms |
| Dimension of input MFCC | 40 | 40 | 40 |
| Dimension of input i-vector | 100 | 100 | 100 |
| Number of hidden layers | 15 | 15 | 15 |
| Architectures of hidden layers | Undisclosed | Undisclosed | Undisclosed |
| Dimension of output PPG vector | 5383 | 5871 | 5996 |
| Training corpus | Undisclosed | Undisclosed | Undisclosed |

Fig. 6 shows the evaluation results of the similarity and naturalness.

Speech synthesized with NKYen_DNN (Approach 1) obtained the highest similarity. As expected, three types of speech that did not use VC (ELN_recording, ELNen_DNN, ELNen_LSTM) had low similarity scores. In other words, they were correctly evaluated to be close to the speaker ELN's voice quality. Focus on two DNN-based methods (NKYen_DNN, NKYen_VC) obtained a higher similarity than three LSTM-based methods (NKYen_GAN_NKYen, NKYen_GAN_NKYja, NKYen_LSTM). The cause is not yet clear, but it can be inferred that DNN, which converts voice quality for each frame, reproduces the speaker's voice quality with higher fidelity than LSTM, which takes the longer context into consideration.

On the other hand, in naturalness evaluation, three types of speech that do not use VC (ELN_recording, ELNen_DNN, ELNen_LSTM) received a high score. In particular, speech synthesized using Bi-LSTM (ELNen_LSTM) achieved a high voice quality approaching the recorded physical voice (ELN_recording), which shows the high quality of Hitachi's speech synthesis. On the other hand, regarding NKY English speech generated with voice quality conversion, DNN-based NKYen_DNN (Approach 1) received the highest evaluation with a 3.28 compared to the 4.18 of the physical voice recording (ELN_recording), realizing a level where productization is possible. In addition, when the conventional methods used in Approach 2 (NKYen_GAN_NKYja) are compared to the improved methods (NKYen_GAN_NKYen), it is clear that the sound quality has greatly improved. The reason that the sound quality of Approach 3 (NKYen_VC) is the lowest is that there is a mismatch of inputted speech, because recorded speech was used in training step and synthesized speech was used in the conversion step.

As a result of considering both similarity and naturalness evaluations comprehensively, we found that the combination of Approach 1 and FF-DNN achieves the highest performance.

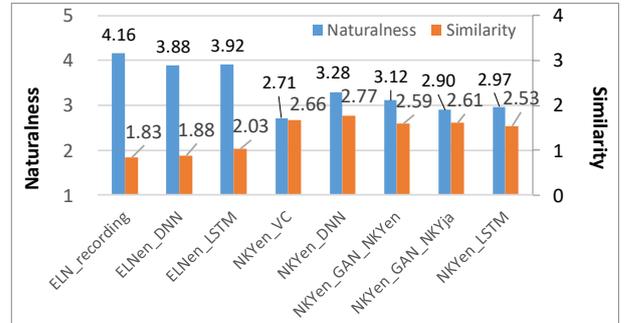

Fig. 6: Results of evaluation test.

## 5. Conclusions & Feature work

In this research, we proposed a new VC method which can convert acoustic and prosodic features same time. Then we tried three approaches to build a multilingual TTS system using the proposed VC method. The results of an evaluation test show that the combination of Approach 1 and FF-DNN achieves the highest performance.

In further work, we will introduce i-vectors or x-vectors to the proposed VC method and build a multilingual TTS system using the customer's own voice.

# 6. References

[1] Y. Qian, H. Liang, and F. K. Soong, "A cross-language state sharing and mapping approach to bilingual (Mandarin–English) TTS," IEEE Transactions on Audio, Speech, and Language Processing, vol. 17, no. 6, pp. 1231–1239, 2009.

[2] F. Xie, F. K. Soong, and H. Li, "A KL divergence and DNN approach to cross-lingual TTS," in IEEE International Conference on Acoustics, Speech and Signal Processing (ICASSP). IEEE, 2016, pp. 5515–5519.

[3] H. Wang, F. Soong, and H. Meng, "A spectral space warping approach to cross-lingual voice transformation in HMM-based TTS," in IEEE International Conference on Acoustics, Speech and Signal Processing (ICASSP). IEEE, 2015, pp. 4874–4878. M. Morise, F. Yokomori, and K. Ozawa, "World: A vocoder-based high-quality speech synthesis system for real-time applications," IEICE Transactions on Information and Systems, vol.E99-D, no.7, pp.1877–1884, 2016.

[4] Y. Stylianou, O. Cappé, and E. Moulines, "Continuous probabilistic transform for voice conversion," IEEE Transactions on Speech and Audio Processing, vol. 6, no. 2, pp. 131–142, 1998.

[5] B. Ramani, M. A. Jeeva, P. Vijayalakshmi, and T. Nagarajan, "A multi-level gmm-based cross-lingual voice conversion using language-specific mixture weights for polyglot synthesis," Circuits, Systems, and Signal Processing, vol. 35, no. 4, pp. 1283–1311, 2016.

[6] M. Abe, K. Shikano, and H. Kuwabara, "Statistical analysis of bilingual speaker's speech for cross-language voice conversion," The Journal of the Acoustical Society of America, vol. 90, no. 1, pp. 76–82, 1991.

[7] M. Mashimo, T. Toda, H. Kawanami, K. Shikano, and N. Campbell, "Cross-language voice conversion evaluation using bilingual databases," IPSJ Journal, vol. 43, no. 7, pp. 2177–2185, 2002.

[8] C.-C. Hsu, H.-T. Hwang, Y.-C. Wu, Y. Tsao, and H.-M. Wang, "Voice conversion from non-parallel corpora using variational auto-encoder," in IEEE APSIPA ASC, 2016, pp. 1–6.

[9] CC Hsu, HT Hwang, YC Wu, Y Tsao, "Voice conversion from unaligned corpora using variational autoencoding wasserstein generative adversarial networks," arXiv:1704.00849.

[10] T. Kaneko and H. Kameoka, "Parallel-data-free voice conversion using cycle-consistent adversarial networks," arXiv:1711.11293, 2017.

[11] H. Kameoka, T. Kaneko, K. Tanaka, and N. Hojo, "Stargan-vc: non-parallel many-to-many voice conversion using star generative adversarial networks," in IEEE SLT, 2018, pp. 266–273.

[12] T. Kaneko, H. Kameoka, K. Tanaka, and N. Hojo, "Cyclegan-vc2: Improved cyclegan-based non-parallel voice conversion," in IEEE Proc. ICASSP, 2019, pp. 6820–6824.

[13] L.-J. Liu, Z.-H. Ling, Y. Jiang, M. Zhou, and L.-R. Dai, "WaveNet vocoder with limited training data for voice conversion," in Proc. Interspeech, 2018, pp. 1983–1987.

[14] L. Sun, H.Wang, S. Kang, K. Li, and H. M. Meng, "Personalized, cross-lingual tts using phonetic posteriorgrams," in Proc. INTERSPEECH, 2016, pp. 322–326.

[15] Y. Zhou, X. Tian, H. Xu, R. K. Das, and H. Li, "Cross-lingual voice conversion with bilingual phonetic posteriorgram and average modeling," in IEEE Proc. ICASSP, 2019, pp. 6790–6794.

[16] T. Yoshimura, K. Tokuda, T. Masuko, T. Kobayashi, T. Kitamura, "Simultaneous modeling of spectrum pitch and duration in HMM-based speech synthesis", Proc. EUROSPEECH, pp. 2347-2350, 1999.

[17] H. Zen, A. Senior and M. Schuster, "Statistical parametric speech synthesis using deep neural networks," Proc. ICASSP, pp. 7962–7966, 2013.

[18] Z.-H. Ling, L. Deng and D. Yu, "Modeling spectral envelopes using restricted Boltzmann machines and deep belief networks for statistical parametric speech synthesis," IEEE Trans. Audio Speech Lang. Process., 21, pp. 2129–2139 (2013).

[19] S. Kang, X. Qian and H. Meng, "Multidistribution deep belief network for speech synthesis," Proc. ICASSP, pp. 8012–8016, 2013.

[20] R. Fernandez, A. Rendel, B. Ramadhadran and R. Hoory, "F0 contour prediction with a deep belief network — Gaussian process hybrid model," Proc. ICASSP, pp. 6885-6889, 2013.

[21] Y. Wang, R. Skerry-Ryan, D. Stanton, Y. Wu, R. J. Weiss, N. Jaitly, Z. Yang, Y. Xiao, Z. Chen, S. Bengio, Q. Le, Y. Agiomyrgiannakis, R. Clark, R. A. Saurous, "Tacotron: Towards End-to-End Speech Synthesis", arXiv:1703.10135, 2017.

[22] J. Shen, R. Pang, R. J. Weiss, M. Schuster, N. Jaitly, Z. Yang, Z. Chen, Y. Zhang, Y. Wang, R. Skerrv-Ryan et al., "Natural tts synthesis by conditioning wavenet on mel spectrogram predictions," in IEEE International Conference on Acoustics, Speech and Signal Processing (ICASSP). IEEE, pp. 4779–4783, 2018.

[23] W. Ping, K. Peng, A. Gibiansky, S. O. Arik, A. Kannan, S. Narang, J. Raiman, J. Miller, "Deep Voice 3: Scaling Text-to-Speech with Convolutional Sequence Learning", arXiv:1710.07654, 2018.

[24] W. Ping, K. Peng, and J. Chen, "Clarinet: Parallel wave generation in end-to-end text-to-speech," in International Conference of Learning Representation (ICLR), 2019.

[25] J. Sotelo, S. Mehri, K. Kumar, J. F. Santos, K. Kastner, A. Courville, and Y. Bengio, "Char2wav: End-to-end speech synthesis," 2017.

[26] Y. Taigman, L. Wolf, A. Polyak, and E. Nachmani, "Voiceloop: Voice fitting and synthesis via a phonological loop," arXiv preprint arXiv:1707.06588, 2017.

[27] Z. Wu, O. Watts, and S. King, "Merlin: An open source neural network speech synthesis system," in 9th ISCA Speech Synthesis Workshop, pp. 218–223, 2016.